\documentclass{article}
\usepackage[utf8]{inputenc}
\usepackage{amsmath}
\usepackage{graphicx}
\usepackage{amssymb}
\usepackage{amsmath}
\usepackage{amsfonts}
\usepackage{authblk}
\usepackage[title]{appendix}
\usepackage{color}
\newcommand{\df}{\mathrm{d}}
\newcommand{\calH}{{\cal{H}}}
\newcommand{\ex}{\mathrm{ex}}            
\newcommand{\co}{\mathrm{co}}            
\newcommand{\bex}{\partial\mathrm{ex}}   
\newcommand{\bco}{\partial\mathrm{co}}   

\newcommand{\vfx}{\mathcal{X}}
\newcommand{\vfy}{\mathcal{Y}} 
\newcommand{\vfz}{\mathcal{Z}}
\newcommand{\vx}{\vfx}
\newcommand{\vy}{\vfy}
\newcommand{\normal}{\mathbf{n}}
\newcommand{\exc}{\nabla\varphi}   
\newcommand{\rot}{J\nabla \psi}  
\newcommand{\hc}{\mathcal{H}_C}    
\newcommand{\hn}{\mathcal{H}_N}    
\newcommand{\hd}{\mathcal{H}_D}    

\newtheorem{theorem}{Theorem}[section]

\newtheorem{definition}{Definition}[section]

\title{Hodge Decomposition of Wall Shear Stress Vector Fields characterizing Biological Flows}
\author[1,2]{Faniry H. Razafindrazaka}
\author[2]{Pavlo Yevtushenko}
\author[1]{Konstantin Poelke}
\author[1]{Konrad Polthier}
\author[2]{Leonid Goubergrits}
\affil[1]{Freie Universität Berlin, Germany}
\affil[2]{Institute for Imaging Science and Computational Modelling in Cardiovascular Medicine, Charit\'{e}-Universit\"{a}tsmedizin Berlin, Germany}
\date{}                     
\begin{document}

\maketitle
\begin{abstract}
A discrete boundary-sensitive Hodge decomposition is proposed as a central tool for the analysis of wall shear stress (WSS) vector fields in aortic blood flows. 
The method is based on novel results for the smooth and discrete Hodge-Morrey-Friedrichs decomposition on manifolds with boundary and subdivides the WSS vector field into five components: gradient (curl-free), co-gradient (divergence-free), and three harmonic fields induced from the boundary, which are called the center, Neumann and Dirichlet fields.
First, an analysis of WSS in several simulated simplified phantom geometries (duct and idealized aorta) was performed in order to understand the impact of the five components. 
It was shown that the decomposition is able to distinguish harmonic blood flow arising from the inlet from harmonic circulations induced by the interior topology of the geometry. Finally, a comparative analysis of 11 patients with coarctation of the aorta (CoA) before and after treatment as well as 10 controls patient was done. 

The study shows a significant difference between the CoA patients and the healthy controls before and after the treatment. 
This means a global difference between aortic shapes of diseased and healthy subjects, thus leading to a new type of WSS-based analysis and classification of pathological and physiological blood flow.
\end{abstract}


\section{Introduction}
Biological flows or hemodynamics of the cardiovascular system play an important role in the genesis, progress and treatment of cardiovascular pathologies including congenital or acquired diseases of the heart, heart valves and vessels. This is because wall remodeling including wall thickness and wall constitution is triggered by hemodynamics. The major hemodynamic parameter describing an interaction between hemodynamics and a vessel wall, which is covered by endothelial cells, is the wall shear stress (WSS). The WSS is an area-normalized tangential force component of the blood flow acting on the wall and/or endothelial cells. 
In turn, endothelial cells trigger and modulate adaptation, inflammation and remodeling of the vessel wall as well as a respective remodeling of the vessel lumen \cite{CHATZIZISIS20072379,Chiu009}. 
Consequently, abnormal WSS is considered an important local risk factor for a set of diseases or pathological processes. 
These include, for example, atherosclerosis of carotid arteries \cite{Zhang2018} or coronary artery disease \cite{SOULIS2010867}, rupture risk of cerebral aneurysms \cite{Meng1924,Boussel2008} or abdominal aortic aneurysms \cite{Stevens2017}, aortic dilatation \cite{Saikrishnan2015}, and thrombus formation \cite{Lozowy2017}. 
Furthermore, the analysis of WSS is also of great interest for the study of the hemodynamic impact of a treatment or a change of the hemodynamics caused by a certain treatment device. 
These studies include, for example, an analysis of post-treatment flow conditions after a treatment of cerebral aneurysms with a flow diverter \cite{JING2016} or a change of flow conditions after an aortic valve replacement \cite{knobelsdorff2014}.
The use of WSS as a reliable biomedical marker characterizing disease, disease progress or initiation and also characterizing hemodynamic outcome of a treatment procedure is challenging. 
This is because WSS is a surface bounded vector field that means that WSS is described by a magnitude and direction varying in space and time. 
This allows for a definition of a set of parameters, which were proposed during the last years as hemodynamic risk parameters for endothelial dysfunction and related wall remodeling. 
A characterization of WSS magnitude, direction, time and space gradients as well as topological features results in a relatively large set of parameters, which are well summarized in \cite{ARZANI2018145}  and \cite{goobergrits2014}. 
The majority of studies investigating WSS in biological flows are numerical studies investigating hemodynamics by an image-based computational fluid dynamics approach  \cite{Morrisheartjnl2015}. 
4D VEC MRI based assessment of the WSS is also proposed in \cite{Rodriguez-Palomares2018}. The primary source of data for the WSS analysis, however, is CFD, since an accurate WSS assessment requires a high spatial resolution as shown by mesh independence studies for CFD solutions \cite{prakash2001}.

Vector fields modelling fluid flow often tend to exhibit a complicated behaviour on various scales and are hard to understand. 
This poses a particular problem for clinical applications where the behaviour of blood  flow in vessels serves as an indicator for potential abnormalities.
The classical Helmholtz decomposition was a first step to classify and analyze vector fields by decomposing them into a divergence-free component and a component having a potential. 
With the advent of Hodge theory, Helmholtz' results generalize to decomposition rules for differential forms on closed manifolds in arbitrary dimensions. 
Since then a tremendous amount of research---both on the theoretical and on the applied side---has been carried out to include manifolds with boundary, differential forms of Sobolev class and various flavours of Hodge-type decomposition statements, 
see e.g. \cite{schwarz-1995} for an overview of Hodge-type decompositions and the survey \cite{bhatia_survey-2013}. 

An important landmark in this evolution is the $L^2$-orthogonal decomposition of $k$-forms on manifolds with boundary as 
\[
\Omega^k = \df \Omega_D^{k-1} \oplus \delta \Omega_N^{k+1} \oplus \df \Omega^{k-1} \cap \delta \Omega^{k+1} \oplus (\calH^k_N + \calH_D^k)
\]
where the spaces $\calH^k_N$ and $\calH_D^k$ of harmonic Neumann and Dirichlet fields, respectively, reflect the absolute and relative cohomology of the manifold. 
Specifically for vector fields, the first two spaces in this decomposition correspond to divergent and rotational irregularities in the interior of the geometry, whereas the latter three spaces represent steady flows through the domain, as each field in these spaces is harmonic. 
A fairly recent result \cite{shonkwiler-2013} provides a further orthogonal
decomposition of these spaces into subspaces  
\[
\calH^k_N = \calH_{N,\co}^k \oplus \calH_{N,\bex}^k \hspace{1em} \text{ and } \hspace{1em} \calH_{D}^k = \calH_{D,\ex}^k \oplus\calH_{D,\bco}^k
\]
which permits a precise distinction between harmonic flows induced by boundary components, represented by the subspaces $\calH_{N,\co}^k$ and $\calH_{D,\ex}^k$, from those induced by the interior topology of the manifold, represented by  $\calH_{N,\bex}^k$ and $\calH_{D,\bco}^k$. 

For the numerical treatment of vector fields it is therefore important to seek for a discretization which on the one hand provides  
a good approximation with predictable error, and on the other hand preserves the structural decomposition results from the smooth theory. 

In this work we focus on a discretization by piecewise constant vector fields
(PCVF) resulting from CFD-based analyses of the blood flow. PCVFs are a very
intuitive and simple to implement approximation while at the same time a concise theoretical framework has been developed in recent years, which includes the aspects of convergence and structural consistency. 
The recent work \cite{Poelke2016126,poelke2017} establishes a consistent discretization for PCVFs of the smooth refined decomposition results for vector fields on surfaces with boundary, now including distinguished subspaces for effective boundary analysis and control. Previous to that, a first strategy for the analysis of vector fields is provided by the decomposition in \cite{polthier_preuss-2003}, with a convergence analysis on closed surfaces in \cite{wardetzky2006thesis}, and a discrete connection for PCVFs is proposed in \cite{azencot2015discrete}, both without an effective boundary control. 

The aim of our study presented here is a proof of concept for the novel Hodge-type decomposition analysis of the WSS vector fields for blood flows in general and specifically for the aortic flow. 
The paper is structured as follows: 
first, a theoretical analysis of each vector field component with respect to a WSS vector field is given.
Second, a detailed description of the  data acquisition and blood flow simulation is exposed.
Finally, a statistical analysis of several patients will summarize the  results.

\subsection{Discrete Hodge-type Decomposition}
The most important results on discrete Hodge-type decompositions on simplicial meshes concerning our application can be summarized by two fundamental theorems: the traditional Hodge-Helmholtz decomposition decomposes vector fields on closed surfaces into three components.  
In contrast, on surfaces with boundary a refined decomposition is provided by the so-called Hodge-Morrey-Friedrichs decomposition. The main ingredients of the discretization and the related spaces are given in the appendix. These decompositions constitute the building block of all analysis in the present
work. For the theoretical foundations see \cite{Poelke2016126,poelke2017}.

\begin{theorem}
[Hodge-Helmholtz decomposition]The space of piecewise constant vector fields
$\Lambda^{1}(M_{h})$ on a closed simplicial surface $M_{h}$ decomposes into an
$L^{2}$-orthogonal sum of the spaces of gradient fields, co-gradient fields
and harmonic fields:%
\begin{align*}
\Lambda^{1}(M_{h})  
& =\nabla S_{h} 
 \oplus J\nabla S_{h}^{\ast} 
 \oplus\left(H:=\ker\operatorname{curl}_{h}^{\ast}\cap\ker\operatorname{div}_{h}\right) \\
\vfx  
& =\underbrace{\nabla \varphi}_{\mathrm{\operatorname{curl}}_{h}^{\ast}\nabla \varphi=0}
 \oplus\underbrace{J\nabla \psi}_{\mathrm{\operatorname{div}}_{h}J\nabla g=0} 
\oplus\underbrace{\vfy}_{\mathrm{\operatorname{curl}}_{h}^{\ast} \vfy = \mathrm{\operatorname{div}}_{h}\vfy=0}%
\end{align*}
\end{theorem}
\begin{figure}[!h]
    \centering
    \includegraphics[width=\linewidth]{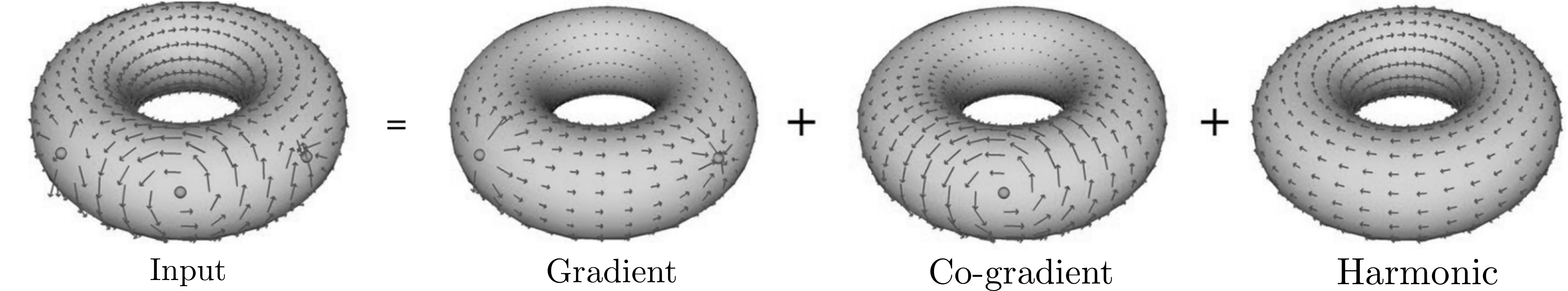}
    \caption{Example of a Hodge-Helmholtz decomposition of a PCVF on a torus into gradient, co-gradient and harmonic field.}
    \label{fig:hh_example}
\end{figure}
The fields belonging to $\nabla S_h$  are free of turbulence and contain only flow induced by sources and sinks. $\rot$ is divergence-free and contains the rotational part of the field (see figure~\ref{fig:hh_example}). Furthermore, if $M_h$ is homeomorphic to a sphere with $m$ boundaries, then the harmonic fields can be  decomposed into three components:

\begin{theorem}[Hodge-Morrey-Friedrichs decomposition HMF]
On a surface $M_h$ homeomorphic to a sphere with $m$ boundaries, the space of harmonic fields can be  decomposed into Neumann fields, center fields, and Dirichlet fields:
\begin{align}
\begin{split}
\label{eq:five_term_decomposition}
\Lambda^{1}(M_{h})  
 &= \nabla S_{0} 
 \oplus J\nabla S_{0}^{\ast}
 \oplus \nabla S_{h}\cap J\nabla S_{h}^{\ast} 
\oplus \rot 
\oplus \hn 
\oplus \hc 
\oplus \hd
\end{split}
\end{align}
\end{theorem}

\begin{figure}[!h]
    \centering
    \includegraphics[width=\linewidth]{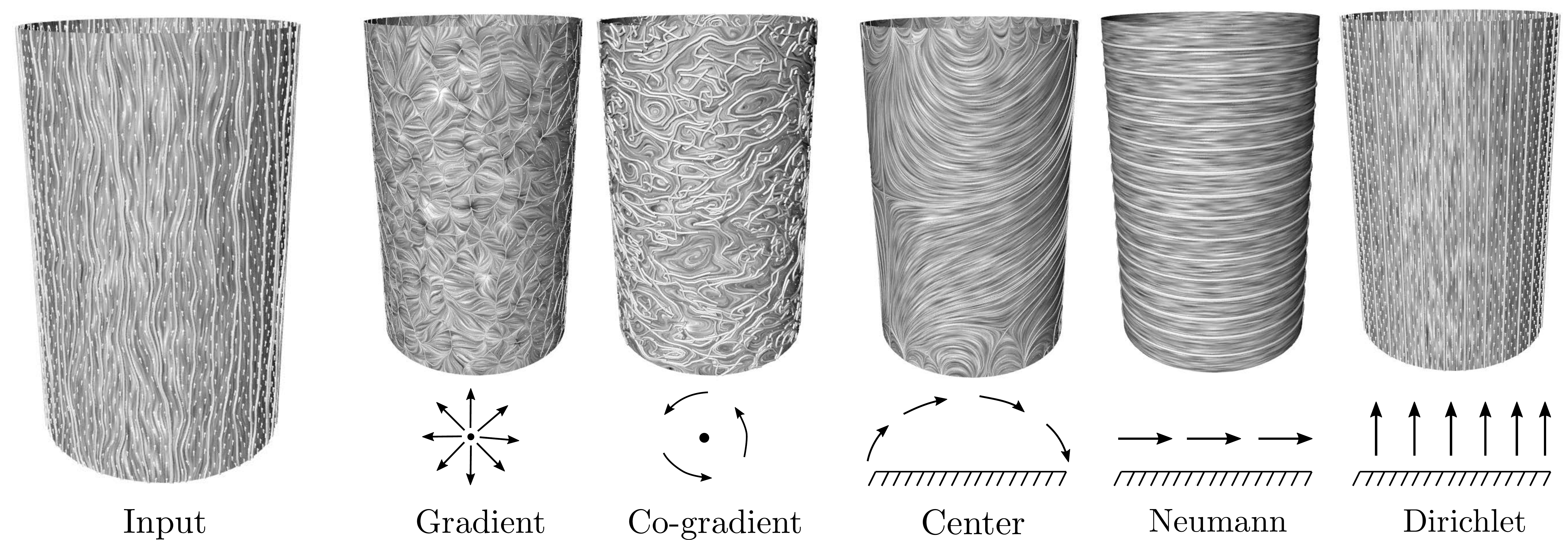}
    \caption{A HMF-decomposition of a perturbed WSS vecor field on a cylinder into five components: gradient, co-gradient, center, Neumann, and Dirichlet vector field.}
    \label{fig:hmf_example}
\end{figure}

One of the main studies of this paper is to understand the nature of these harmonic spaces on  simulated CFD WSS vector fields. 
Intuitively the space $\nabla S_{h}\cap J\nabla S_{h}^{\ast} $ of center vector fields behaves similarly to the space of smooth vector fields forming an $\approx45^{\circ}$ angle with the boundaries, the Neumann vector fields are orthogonal to the boundaries, and the Dirichlet are parallel to the boundaries. 
By the Pythagorian theorem it is 
\[
\|\vfx\|^2 = \|\exc\|^2+\|\rot\|^2+\|\hn\|^2+\|\hc\|^2+\|\hd\|^2 
\]
which enables a full quantification of the input vector fields according to their decomposition components. 
Figure~\ref{fig:hmf_example} shows an example of a HMF-decomposition on the WSS of a simple flow on a cylinder. Notice how the field is dominated by $\hd$. 

\subsection{WSS Component Analysis}
In this section, we study each component of the HMF-decomposition with respect to the WSS of several phantom as well as real patient models   obtained from CFD. The phantom models are either hand-designed or real patient models with mathematical deformation and boundary conditions. The observations are used to emphasize on possible changes of WSS encoded in each HMF-components with respect to anatomy/topology of the geometry, and parameters used for blood flow simulation.

\subsubsection{Perturbed WSS}
\begin{figure}[!h]
    \centering
    \includegraphics[width=\linewidth]{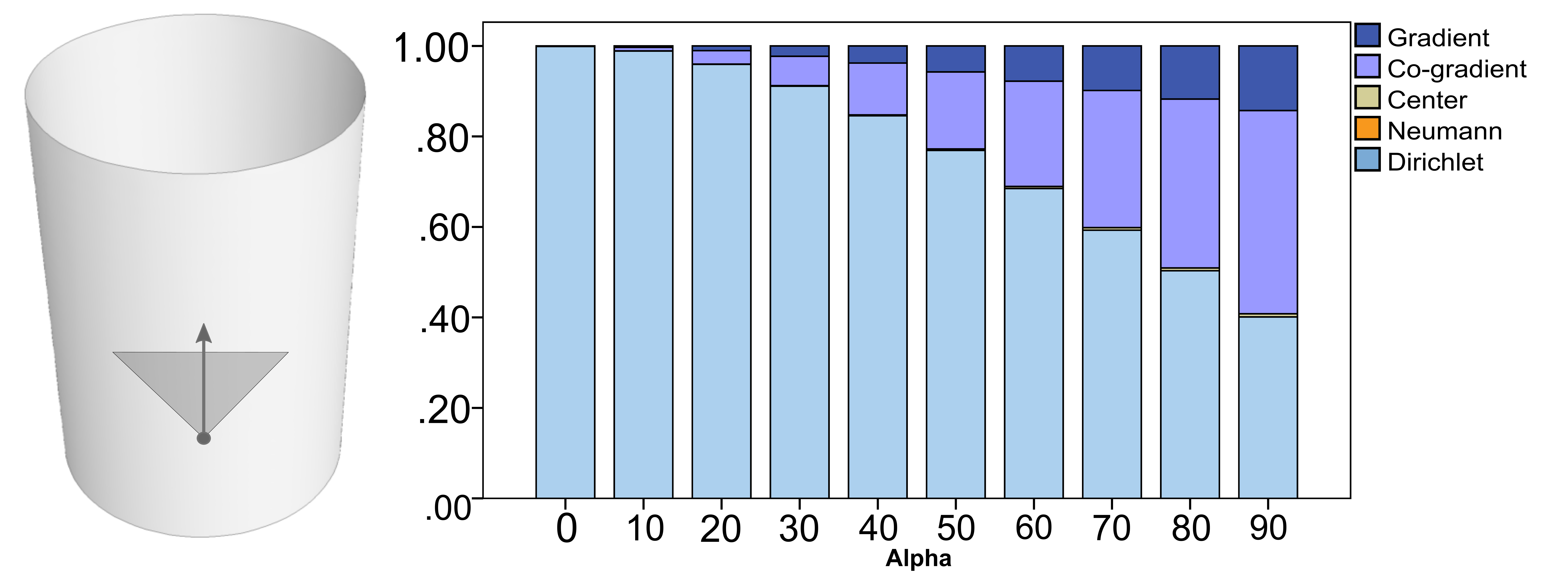}
    \caption{Perturbation of a laminar WSS on a cylinder starting from $0^{\circ}$
to $90^{\circ}$. An increase in angle deviation decreases the Dirichlet component
and increases the co-gradient components.}
    \label{fig:diagram0}
\end{figure}

Consider a smooth cylinder with  a WSS of a laminar flow. We add a moderate amount of rotational noise to the vector field within the interval $(-\alpha,\alpha)$ where $\alpha$ bounds the frequency of the noise. High values of $\alpha$ correspond to high overall frequencies while small values alter slightly the global smoothness of the flow. The HMF-decomposition shows that the Dirichlet field $\hd$ recovers the original field in its unperturbed state,  behaving similarly to a vector field denoising. The increase of $\alpha$ decreases $\hd$ and increases the co-gradient field.  Figure~\ref{fig:diagram0} is a quantitative comparison of each decomposition where $\alpha$ varies from $0^{\circ}$ to $90^{\circ}$ degree. The diagram shows that $\hd$ is a good reference to understand  the global structure of the WSS. In general, harmonic fields depend only on the topology of the shape, not the field. In figure~\ref{fig:centerNeumann} (second row), for example, $\hd$ stays invariant even though the input velocity profile is changed. Quantitatively half of the WSS component is Dirichlet. One logic behind this is reflected in the nature of fluids, being mostly dominated by a laminar component in order to move only in one direction.  

\subsubsection{Coarctation analysis}

\begin{figure}[!h]
    \centering
    \includegraphics[width=\linewidth]{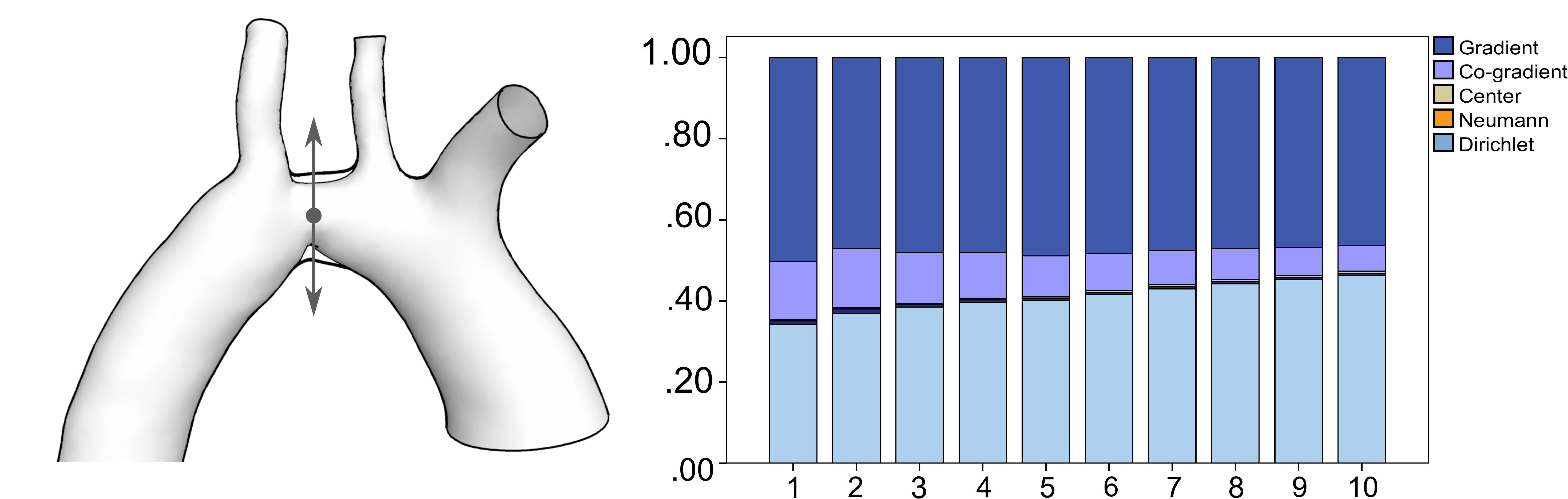}
    \caption{Linear deformation of a pre  to post intervention of a patient
stenosis. A constant input velocity profile is used for the simulation. An increase in the Dirichlet component and  a reduction in the co-gradient
field  is observed within 10 frames of the deformation.}
    \label{fig:deformation}
\end{figure}

Aortic coarctation is a common congenital heart disease. It represents a local
narrowing of the aortic vessel causing abnormal blood flow and pressure and in the
cardiovascular system. Generally, the WSS vector field  of a pre- and post-operative
patient does not provide enough information about amelioration in the patient blood
flow. The HMF-decomposition enables us in a theoretical setting to identify
important changes between the two states. We took a segmented MRI scan of a patient
before and after operation, deformed the coarctation linearly from pre to post and
analyzed the WSS evolution during the diffusion process. The simulation is performed
with a plug profile and settings given in section~\ref{sec:input}. The results are
shown in Figure~\ref{fig:deformation}. We notice a significant increase in the
Dirichlet field amortized with a reduction in  co-gradient field. The improvement in
the Dirichlet field component corresponds to the improvement of the overall blood
flow as proven previously. Notice how the gradient, Neumann, and center fields
remain almost unchanged. The nature of these components is explained in the next
sections.

\subsubsection{Plug vs MRI profile}
\begin{figure}[!h]
    \centering
    \includegraphics[width=\linewidth]{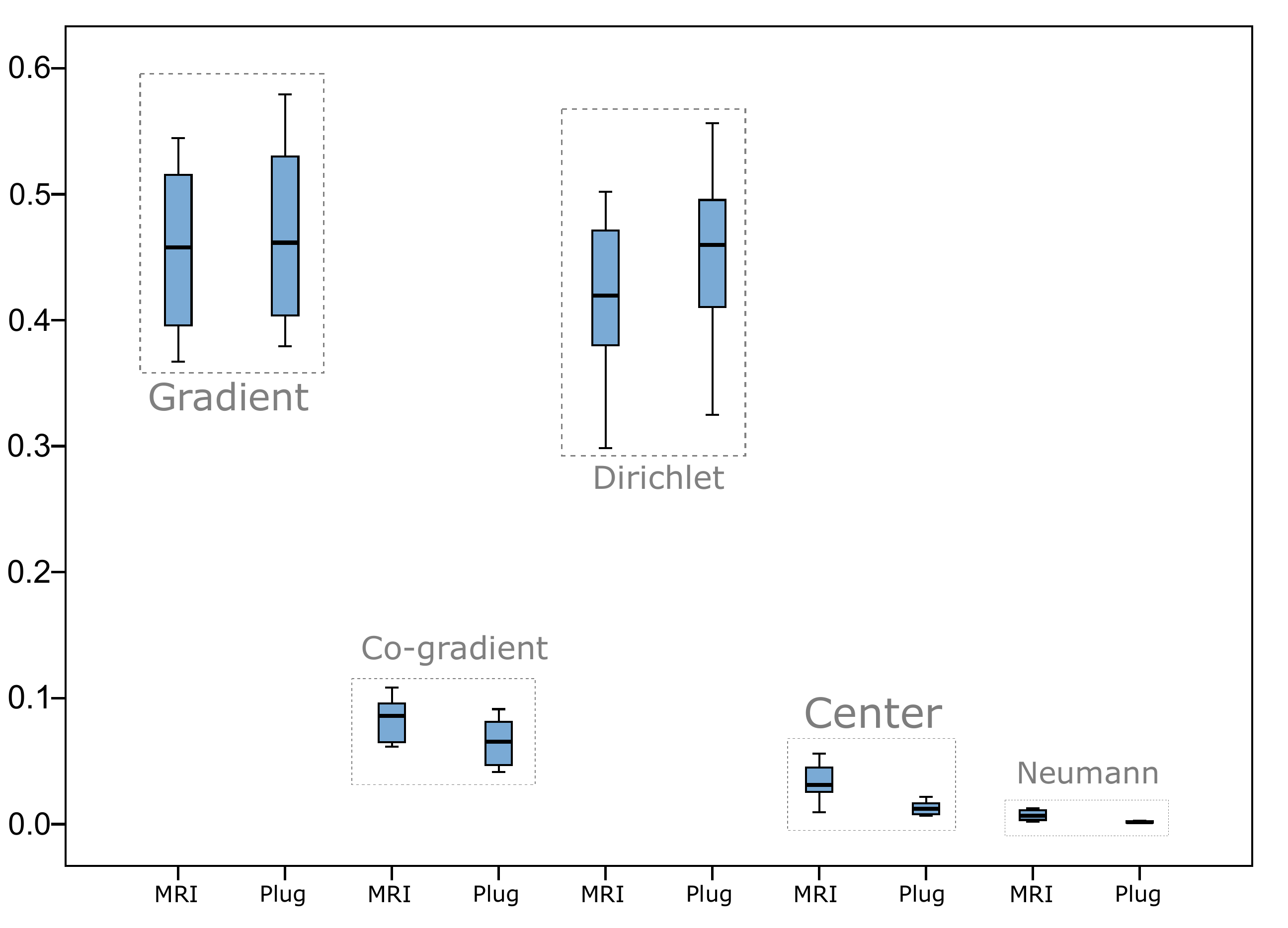}
    \caption{Comparison of the WSS of  10 healthy patients with MRI vs Plug inlet velocity profile. The noise produced from the MRI can be identified by a significant increase of center and Neumann fields.}
    \label{fig:boxplotMRIvsPlug}
\end{figure}

The boundary conditions used in CFD are, most of the time, either a constant input
velocity field (plug) or velocity information acquired from 4D MRI scans using
specialized software and sequences. MRI profiles are noisy, however, and sometimes
the resulting WSS field looks more perturbed than a WSS field obtained by a plug
profile. Using the HMF-decomposition, one can classify which components of the WSS
are more affected by the inlet velocity profile. . 
\begin{figure}[!h]
    \centering
    \includegraphics[width=.6\linewidth]{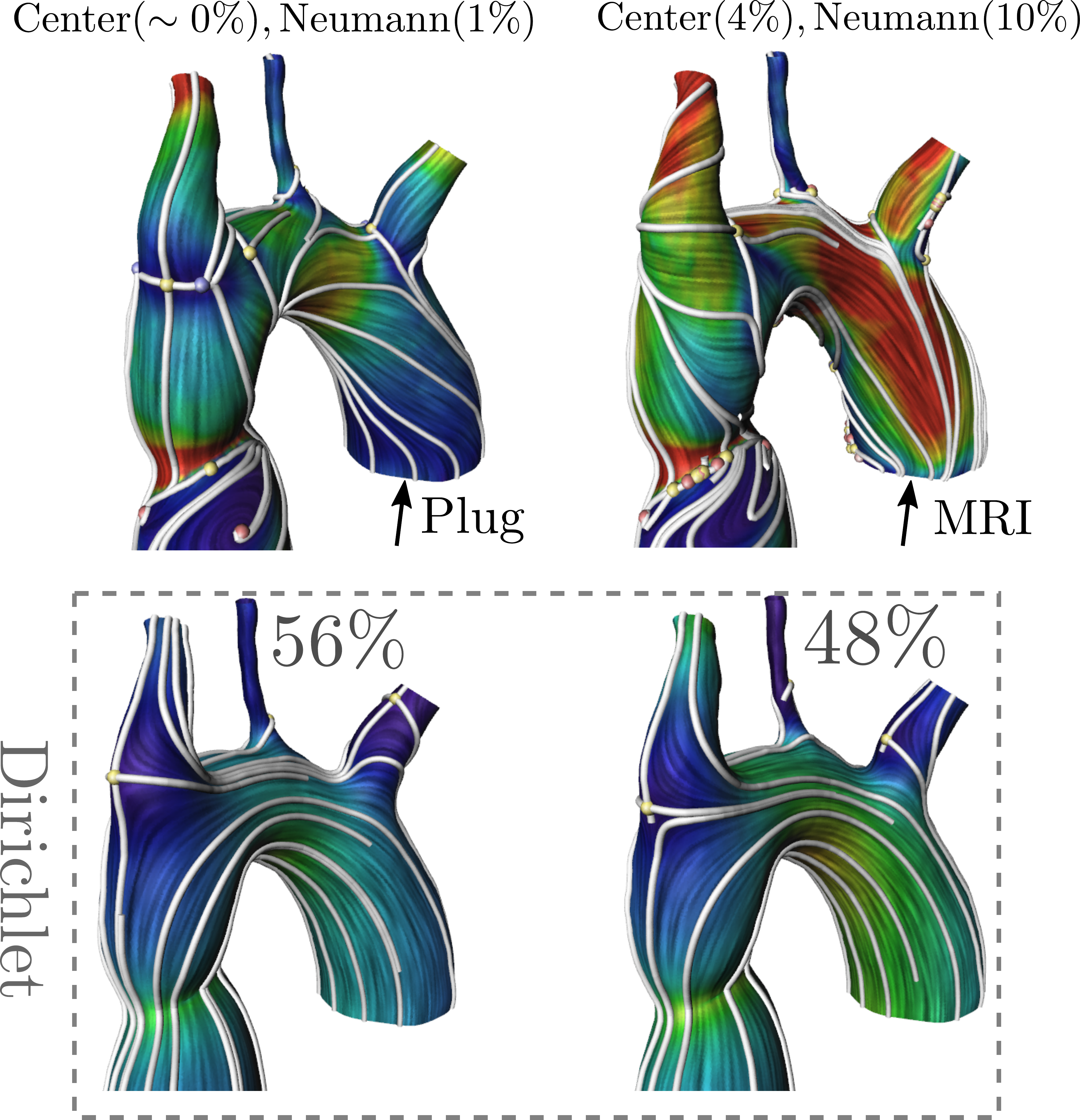}
    \caption{First row: Plug vs MRI input profile encoded in the center and Neumann components. Second row: invariance of the Dirichlet component under the change of input profile.}
    \label{fig:centerNeumann}
\end{figure}
Figure~\ref{fig:centerNeumann} presents a WSS analysis of the same patient with a
different inlet profile. The two vector fields are very different but by analyzing
each HMF-decomposition component, one can see perturbations in the center and
Neumann fields. The Dirichlet fields in both cases are topologically the same
(similar streamline and same number of singularities). 
+Figure~\ref{fig:boxplotMRIvsPlug} is a comparison of both inlet boundary conditions
for ten control patients. The statistical analysis (paired t-student test) of
normally distributed data (Komogorov-Smirnov test) found no significant difference
for the gradient component (p=0.571). 
+However, significantly (p=0.038) smaller Dirichlet components for MRI-measured
inlet velocity profiles accompanied with significantly larger co-gradient (p=0.007),
center (p=0.002) and Neumann (p=0.002) components of the HMF-decomposition have been
observed. Notice that the $L^2$-norms of the center and Neumann components in both
cases are relatively  small compared to the other components.

\subsubsection{Number of Branches}
The following study shows the effect of branches on an idealized aorta. Starting with a curved cylinder with zero branch, artificial branches are
successively added and a blood flow is simulated on each geometry using a plug inlet
profile. The results show that for this ideal situation the gradient field increases
with the number of branches. Geometrically, branches induce a high curvature and hence more divergence. There
are still several parameters not taken into account such as tapering or twisted
cylinders. The proposed setup with the correct geometry can be used to analyze these
extra cases.  
\begin{figure}[!h]
    \centering
    \includegraphics[width=\linewidth]{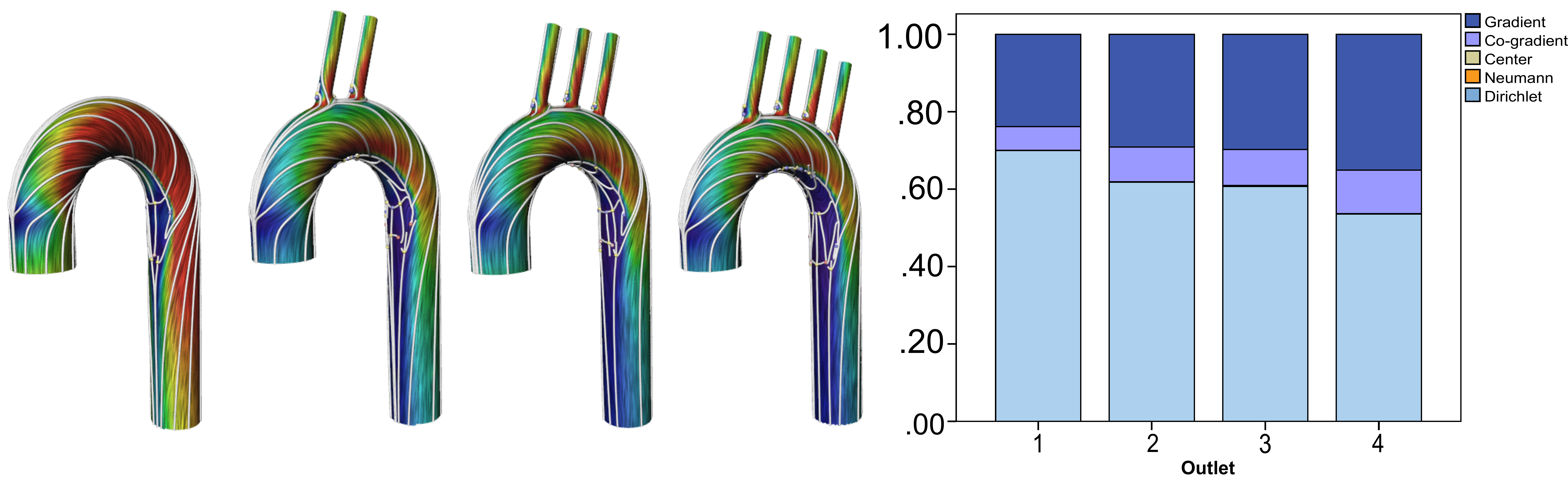}
    \caption{Analysis of artificial aorta models with a varying number of outlets. 
    	The second row shows the flow of the gradient field. 
    	The diagram shows that the number of branching outlets is closely related to the gradient and Dirichlet field.}
    \label{fig:aorta}
    \end{figure}

\subsubsection{Unsteady flow}
\begin{figure}
\centering
    \begin{tabular}{c}
    \includegraphics[width=.9\linewidth]{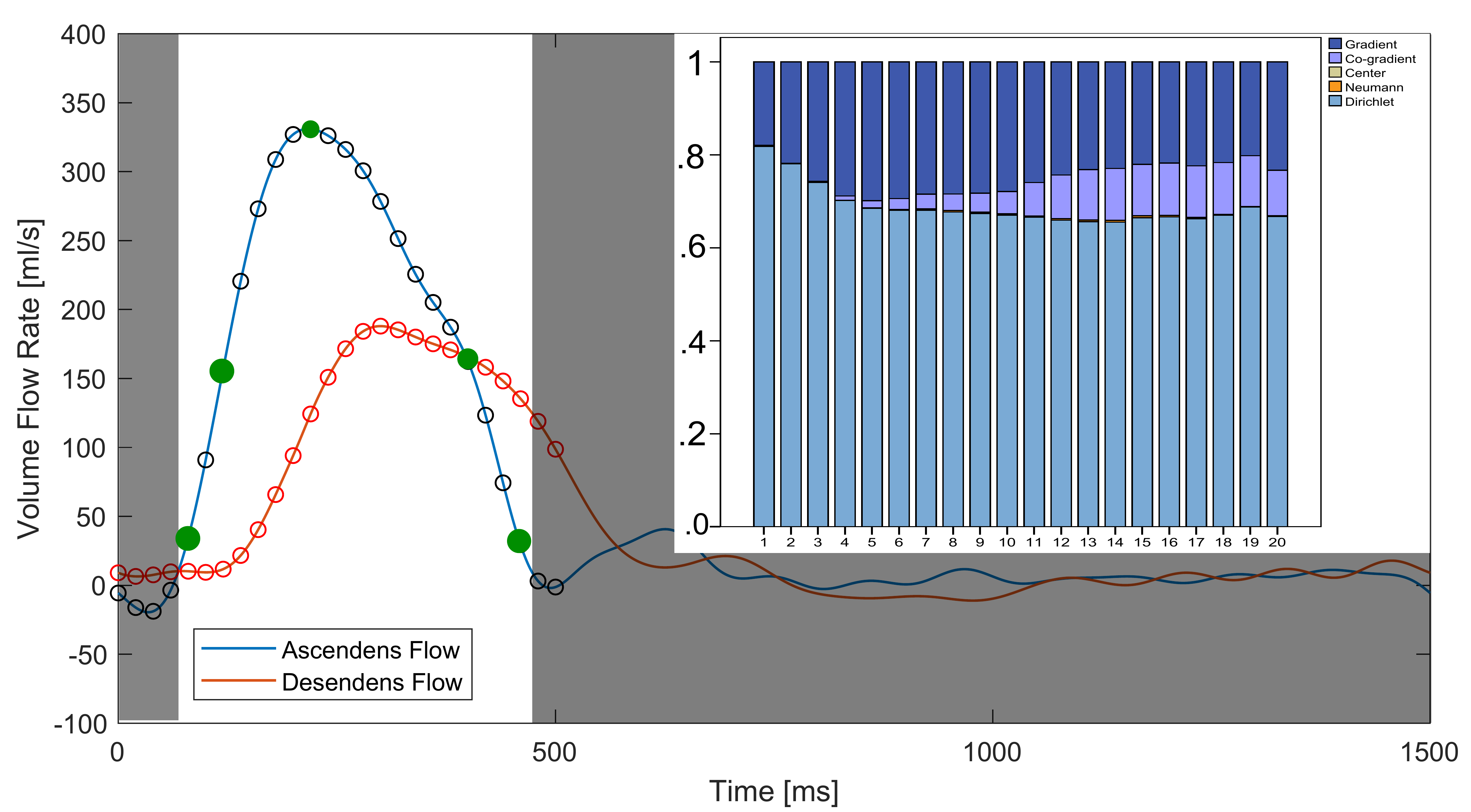}\\
    \includegraphics[width=.9\linewidth]{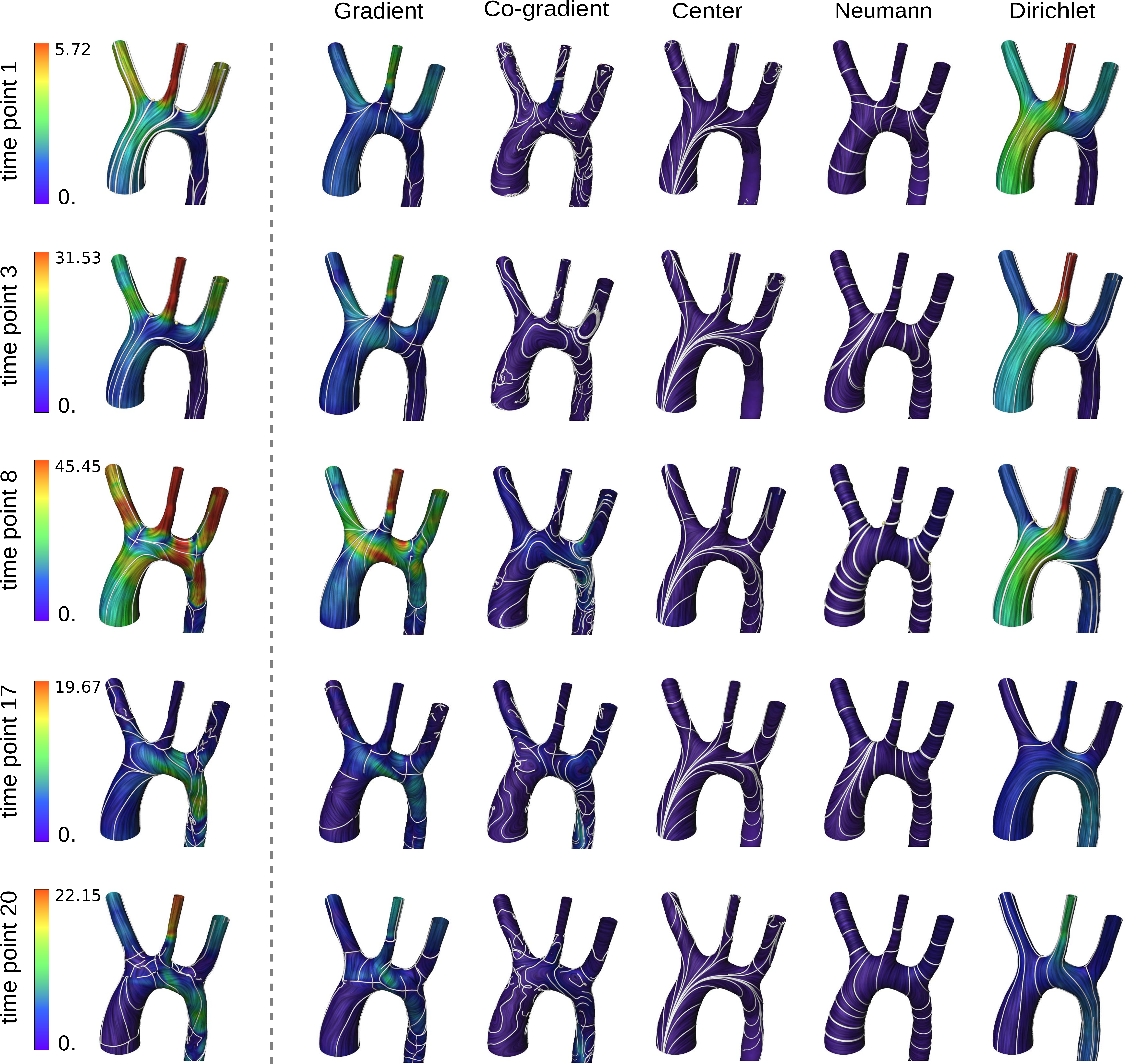}
    \end{tabular}
    \caption{Evolution of the WSS HMF-decomposition in a CFD simulation with an
unsteady flow. The diagram shows 20 time points of the simulation. The close-ups are
five phases from the 20 time points of the unsteady flow simulation (green dots).
The colors are relative to the min-max  magnitude of each input vector field. }
    \label{fig:unsteady_diag}
\end{figure}

Finally, the effect of time-varying flow boundary-conditions is examined in this section. For this purpose, an unsteady CFD simulation of a whole cardiac cycle performed earlier for a MICCAI CFD Challenge~\cite{Schaller2014} was used. For the analysis however, only the systolic part of the cardiac cycle is considered, since the diastolic part shows only little to zero flow and therefore negligible WSS in the aorta. Twenty time-points have been evaluated in total and are presented in figure \ref{fig:unsteady_diag}, along with inlet and outlet flow-curves. Additionally, decomposed WSS vector field plots are presented for five time points with a more detailed picture of the WSS distribution.
As can be seen from the decomposition at the various time points, the respective components of the HMF-decomposition change over time, with a significant increase in co-gradient component and decrease in Dirichlet component. Furthermore, it appears that the variations of the HMF-components do not only arise from variations of the flow rate (i.e. Reynolds Number) but also from acceleration and deceleration effects. This can be seen by comparing two timepoints with equivalent flow rates, as for example timepoints 3 and 17, both of which show a flow rate of about 160 ml/s. Despite that, timepoint 17, at which the flow is being decelerated, shows a significantly higher co-gradient and lower gradient component than timepoint 3, where the flow is being accelerated. Neumann and center components however remain at almost zero throughout the whole systole.

\section{Method}
\begin{figure}[!h]
    \centering
    \includegraphics[width=\linewidth]{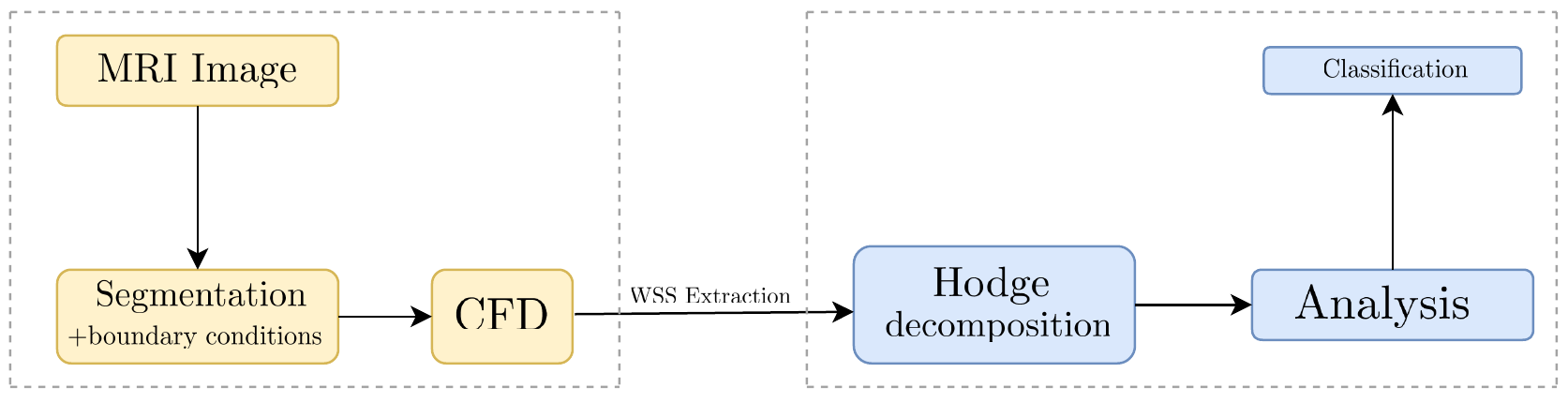}
    \caption{Analysis pipeline of WSS vector fields extracted from a simulated
model and analyzed via Hodge decomposition.}
    \label{fig:diagram}
\end{figure}

A diagram summarizing the analysis pipeline is given in figure~\ref{fig:diagram}.
The implementation of the HMF-decomposition is done following the  iterative
$L^2$-projection approach~\cite{Poelke2016126,polthier_preuss-2003} with the
discretization given in the appendix. The choice of basis functions for each harmonic field subspace
follows~\cite{Poelke2016126}. Our system takes as input a mesh with a vector field and return the five components decomposition of the vector field, assuming that the surface fulfills the topological requirement. Our system takes as input a mesh with a vector field and returns the five-term
decomposition~eq.~\eqref{eq:five_term_decomposition} of the vector field, assuming
that the surface fulfills the topological requirement. Our implementation is done in
Java using the JavaView (www.javaview.de) geometry processing package. The line
integral convolution (LIC) implemented in ZIBAmira 2015.28 (Zuse Institute Berlin)
is used for the field visualization.  Maximum magnitude is colored with red while
close to zero vectors are colored in violet. Most of the data used in this paper is
from  real patient biological models..

\subsection{Data Input}
\label{sec:input}
\paragraph{MRI:}
The HMF-decomposition analysis was done for WSS vector fields of the aorta from a MRI based CFD analysis of the aortic flow. These are subdivided in two groups: controls and coarctation of the aorta (CoA) patients before and after treatment. 

The study was carried out according to the principles of the Declaration of Helsinki and approved by the local ethics committee. Written informed consent was obtained from the participants and/or their legal guardians.

MRI examinations used to set boundary conditions for the CFD analysis were
performed using a 1.5 Tesla Achieva R5.1.8 MRI scanner with a five-element cardiac
phased-array coil (Philips Medical Systems, Best, The Netherlands). MRI protocols
including a  routine three-dimensional anatomical imaging in end-diastole are used
to reconstruct the geometry of the aorta (3D MRI). The sequence parameters used
were: acquired voxel size $0.66\times0.66\times3.2$ mm, reconstructed voxel size
$0.66\times0.66\times1.6$ mm, repetition time 4 ms, echo time 2 ms, flip angle
$90^{\circ}$, number of signal averages 3. Four-dimensional velocity-encoded MRI (4D
VEC MRI) was used to capture the flow data of the ascending aorta and the thoracic
aorta (acquired voxel size $2.5\times2.5 \times2.5$ mm, reconstructed voxel size
$1.7\times1.7\times2.5$ mm, repetition time 3.5 ms, echo time 2.2 ms, flip angle
$5^{\circ}$, 25 reconstructed cardiac phases, number of signal averages 1). High
velocity encoding (3-6 m/s) in all three directions was used in order to avoid phase
wraps in the presence of valve stenosis or secondary flow. All flow measurements
were completed with automatic correction of concomitant phase errors. These data
were used to set inflow and outflow boundary conditions. 

\begin{figure}[!h]
    \centering
    \includegraphics[width=\linewidth]{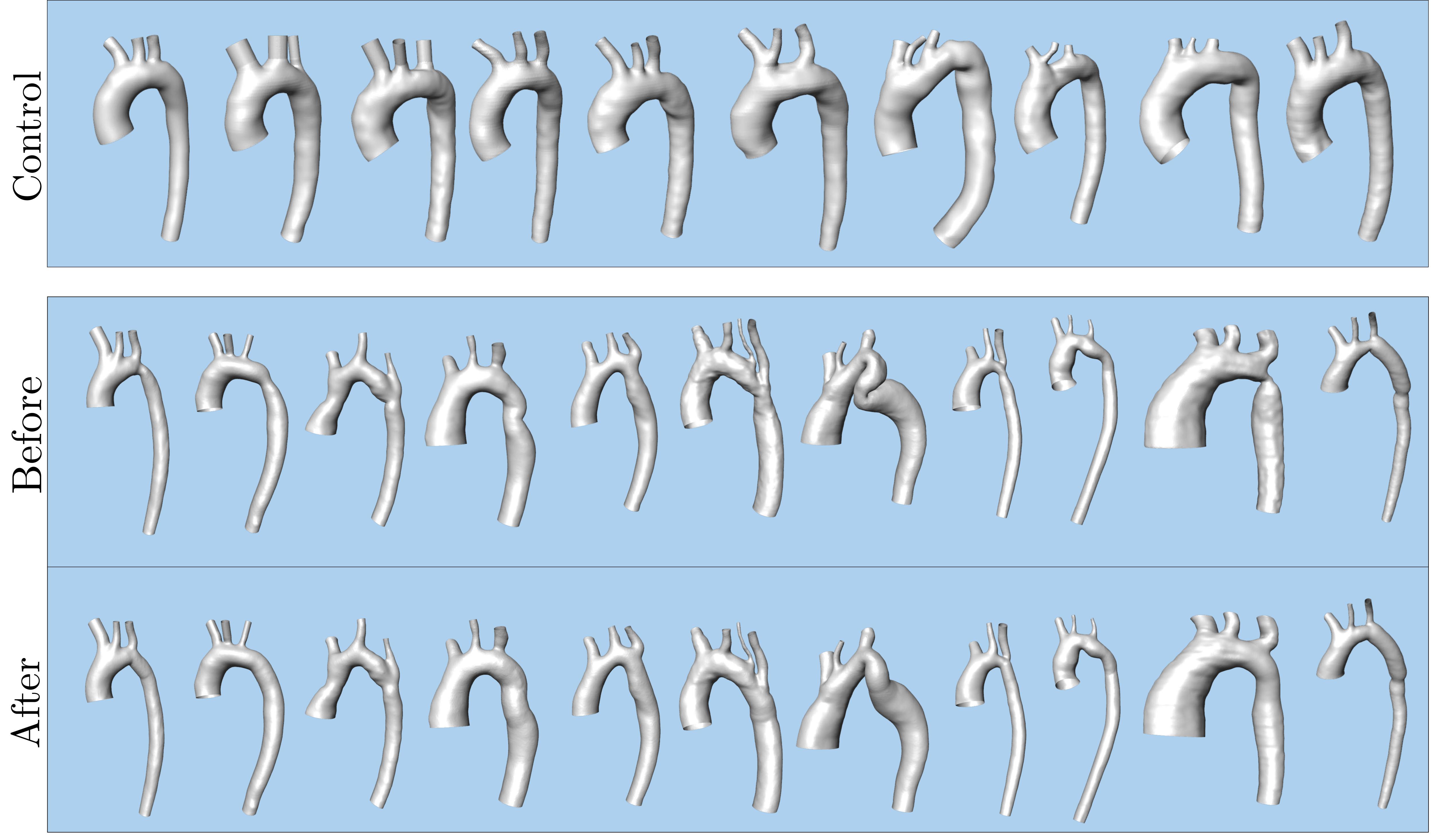}
    \caption{Segmented aorta reconstructed from MRI images and used for the CFD simulation.}
    \label{fig:table}
\end{figure}

\paragraph{CFD:} CFD requires geometries. Geometries of human aortas were segmented
and reconstructed using ZIBAmira 2015.28 (Zuse Institute Berlin, Berlin, Germany) 
according to the previous description \cite{florian2018}. Briefly, intensity based
image segmentation was done semi-automatically with an intense manual interaction.
Rough surface geometries were then generated from segmentations with a subvoxel
accuracy and subsequently smoothed using Meshmixer (v. 3.3, Autodesk, Inc., San
Rafael, USA). These procedures were described in more detail
earlier~\cite{florian2018}. Figure~\ref{fig:table} shows all aorta models used for
our analysis.

With the exception of the unsteady case, all simulations were performed as steady-state simulations of the peak-systolic aortic flow using STAR-CCM+ (v. 12.06, Siemens PLM Software, Plano, USA). Vessel walls were assumed to be rigid and a no-slip boundary condition was applied at all walls. To model turbulence observed in systolic aortic hemodynamics, a $k-\omega$ SST turbulence model with a turbulence intensity of 5 percent at the velocity inlet was used. Blood was modelled as a non-Newtonian fluid with a constant density of $1050\,\,kg/m^3$ and a Carreau-Yasuda viscosity model~\cite{KARIMI201442}. Patient-specific flow rates as
measured with GTFlow (GyroTools LLC, Zurich, Switzerland) from 4D VEC MRI data were
set at the LVOT inlet and the descending aorta outlet. Furthermore, patient-specific
velocity profiles at peak systolic flow rate were extracted using MEVISFlow (v.
10.3, Fraunhofer MEVIS, Bremen, Germany) and set as inlet boundary conditions. The
used CFD pipeline was earlier validated by a comparison with 4D VEC MRI measured
velocity fields as well as clinically validated against catheter measured pressure
drops in cases of CoA~\cite{goubergrits2015}. Furthermore, to validate results of
our simulations we compare velocity fields calculated by CFD against velocity fields
measured by 4D flow MRI, both visualized by velocity magnitude color coded path
lines~\cite{goubergrits2013}. Calculated wall shear stress values are in the range
of published results~\cite{ladisa2011}.

\subsection{Statistical analysis}
Statistical analysis of the Hodge Decomposition results was done using the software package IBM SPSS Statistic, version 25 (IBM, USA). Measured data are presented as mean and standard deviation (SD) for normally distributed data or as a median with IQR. All data were tested for normality using the Kolmogorov-Smirnov-Test. 
Depending on the results of the normality test, the T-student test or
Mann-Whitney-U test were used for the group comparison. Paired tests were used to
compare pre- and post-treatment results. A $p$ value $< 0.05$ was considered
significant.

\section{Results}



\begin{figure}[!h]
    \centering
    \includegraphics[width=\linewidth]{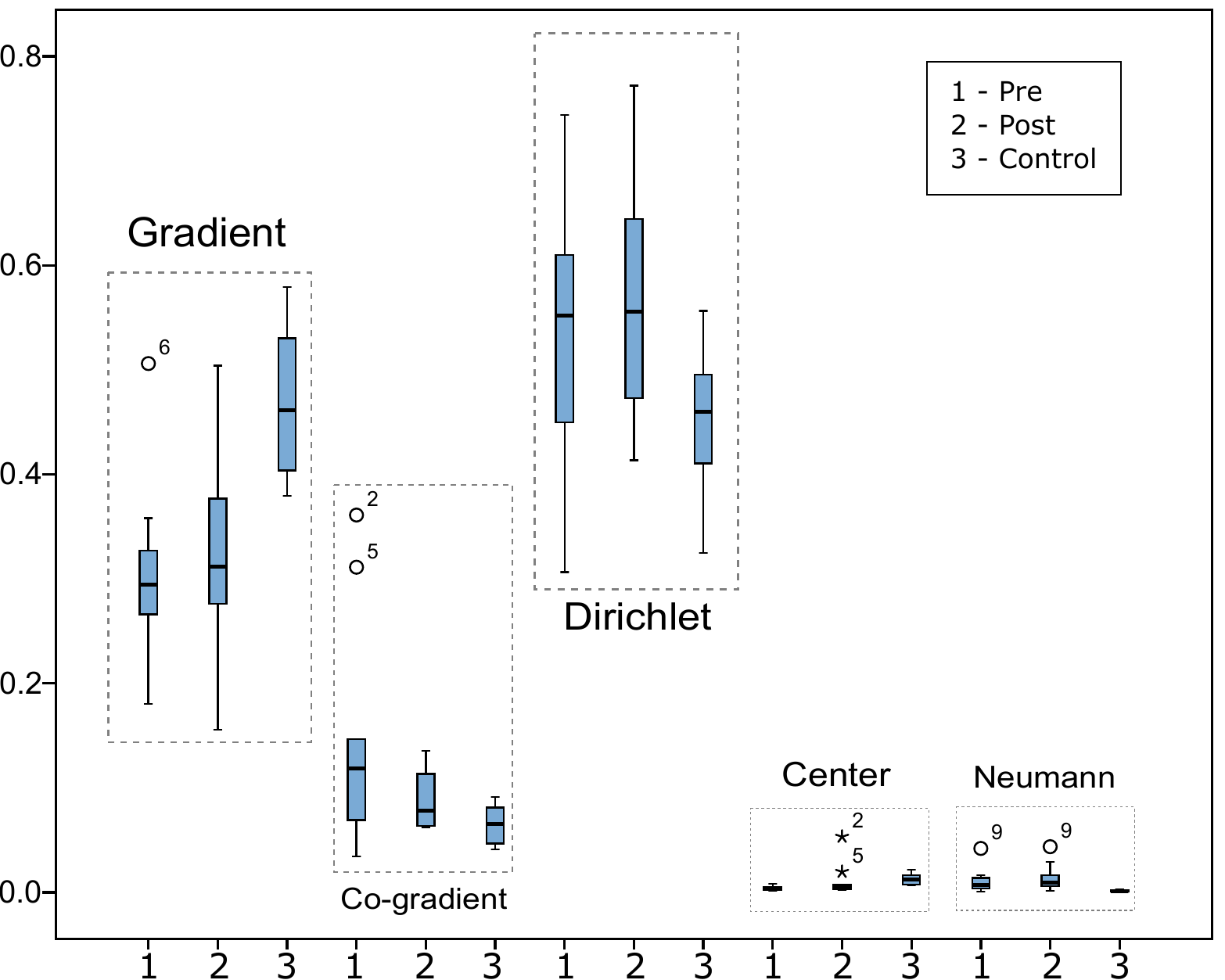}
    \caption{Comparison of the WSS of 11 patients before and after intervention,
and 10 healthy patients. An improvement in gradient and Dirichlet together with a
reduction in co-gradient is observed.}
    \label{fig:boxplot}
\end{figure}

Results of the HMF-decomposition analysis of 11 CoA patients before and after treatment as well as 10 controls are illustrated  in figure~\ref{fig:boxplot}.  The T-student test found significantly lower gradient and significantly higher Dirichlet in CoA cases before treatment vs. controls: 0.3 (SD=0.083) vs. 0.46 (SD=0.065) gradient, and 0.54 (SD=0.125) vs. 0.42 (SD=0.065) Dirichlet. The co-gradient in the CoA group was higher as in controls with median 0.119 IQR [0.069-0.147] vs. median 0.086 IQR [0.065-0.098], approaching significance (Mann-Whitney test, p=0.061).
Overall significant reduction (paired Wilcoxon test, p=0.041) in co-gradient has been observed from pre (median 0.119 IQR [0.069-0.147]) to post intervention (median 0.070 IQR [0.064-0.113]) as expected from the theoretical experimentation  exposed previously. 
However, no significant changes in the major flow descriptors of gradient (p=0.174) and Dirichlet (p=0.073) were found between pre and post treatment WSS vector fields (paired T-Student test).

\section{Discussion}
Our first results on the application of a discrete HMF-decomposition analysis of the aortic flow and especially an analysis of the WSS vector fields of the coarctation of the aorta (congenital narrowing of the aorta) disease revealed great potential for computational biofluid mechanics. Based on the results shown in figure~\ref{fig:boxplot} we suppose that the HMF-decomposition analysis allows us to find anatomical shapes forming pathological hemodynamics before disease progress becomes symptomatic. 

Our findings show an added value of the HMF-decomposition analysis if compared with
the usually used analysis of WSS vector fields by visualization or quantification of
time- and surface-averaged WSS values, areas with low WSS values (e.g. WSS values
below 0.5 Pa) or areas with high OSI as well as an analysis of WSS critical
points~\cite{Boussel2008,ladisa2011,ARZANI2018145}. This approach, however, does not
allow, for example, a quantitative analysis of two different abnormal WSS vector
fields or a quantitative analysis of different impact factors (boundary conditions)
forming abnormal hemodynamics.

The results shown in figure~\ref{fig:boxplot} together with the theoretical
analysis on ideal models raise several open questions. Could  a pathological anomaly
such us stenosis present in the aorta  be identified by its amount of WSS
co-gradient? Control healthy patients have less co-gradient field. The pre vs post
operative patient also show a significant improvement in co-gradient field. An
objective classification has not been achieved with our current analysis because of
the limited number of patient models. Nevertheless the theoretical deformation shown
in figure~\ref{fig:deformation} suggests that it should generally be the case.

The current analysis is focusing only on the studying the differences in WSS vector fields shown by the HMF-decomposition due to treatment aiming at restoring the stenosed region towards a physiological diameter. The differences between diseased and control groups aiming to identify hemodynamic reasons for the development of a pathological anatomy are emphasized. Future research could be also focused on the impact or decomposition of hemodynamic and/or morphometric boundary conditions on the resulting HMF WSS vector field decomposition. This is, however, a challenging task since the hemodynamics depend on a set of non-linear effects of all boundary conditions including the flow rates distributions, vessel curvature, branching topology and others.

A perfect flow would have a pure WSS Dirichlet field, but due to branches and
taperings in the shape, gradient and co-gradient components are also present. On the
one hand, there are higher gradient than Dirichlet components in the control group,
on the other hand there are higher Dirichlet than gradient components in the pre and
post operative groups.  The theoretical analysis on the  number of branches shows
that the nature of the gradient field may change and become dominant. Understanding
the correlation between the gradient and the Dirichlet field will be a good
direction for  future research.  

We applied the HMF-decomposition first to analyse WSS vector fields, since WSS is a known risk factor for the genesis and progress of pathological processes associated with an interaction between blood flow and vessel wall.
The analysis allows for an integral characterization of the WSS distribution. 
However, it does not replace an analysis of WSS magnitudes, which are also associated with abnormal blood flow conditions: 
regions with low WSS promote development of atherosclerosis and thrombus
formations, whereas high WSS could cause an injury of endothelial cells.  As part of our study we investigated the impact of side branches, degree of
stenosis and/or treatment procedure, inlet flow profile boundary conditions and the
impact of laminar flow disturbances on WSS vector fields as characterized by the
HMF-decomposition.

The HMF-decomposition analysis of simulated WSS vector fields was based on the
Reynolds-averaged Navier-Stokes (RANS) solver using the $k-\omega$ SST turbulence
model. However, flow simulations of hemodynamics allowing assessment of pressure and
velocity fields and hence WSS are not limited to the RANS CFD. The Lattice-Boltzman
method (LBM), Large-Eddy simulations (LES) or Smoothed-Particle Hydrodynamics (SPH)
are possible CFD alternatives. For example, LES is supposed to be better suited in order to simulate accurately
transition to turbulence and to assess turbulent structures~\cite{vergara2017}.
Finally, the choice of the CFD approach should be done based on validation studies
comparing simulation results vs. in vivo measurements~\cite{miyazaki2017}. The
HMF-decomposition analysis is, however, independent from the CFD approach.

Further possible and planned studies include, for example, an analysis of pulsatile flows, analysis of flow differences due to different turbulence models, the extension of an analysis to other parts of circulation (e.g. coronary arteries, carotid bifurcations or cerebral vessels) and other diseases (e.g. abdominal aortic aneurysms, cerebral aneurysms or coronary artery disease).

Summarizing our results, the HMF-decomposition is able to support 
(1) basic research of the flow mediated disease, 
(2) predictive computational modelling of the treatment procedure as well as 
(3) quantitative analysis of the hemodynamic treatment outcome. 
Altogether, it supports a clinical translation of the computational modelling approach.

\section{Conclusion}
The novel discrete Hodge-Morrey-Friedrichs decomposition was for the first time
applied to analyze the WSS vector fields of simulated patient-specific aortic blood
flows.  The approach seems to be a powerful tool to distinguish between pathological and
physiologic blood flows, 
+and to characterize the impact of inflow boundary conditions as well as the impact
of a treatment. 


\begin{appendices}
\label{sec:appendix}
\section{}

In this appendix, we give a brief introduction to the calculus on discrete
surfaces. Only the most relevant notions necessary to understand the discrete Hodge
decomposition are given. A complete overview can be found
in~\cite{polthier_preuss-2003}.

\subsection{Simplicial Surfaces}
A 2-dimensional \textit{simplicial surface} $M_h$ is a set of triangles glued at their edges with a manifold structure. In finite element analysis, this type of discrete surface is called \textit{triangle mesh}. 
For actual FEM computations on such meshes one often uses the space $S_h$ of linear Lagrange functions, or the space $S^{*}_h$ of Crouzeix-Raviart functions. 
They are defined by
\begin{align*}
S_{h}  &  :=\left\{  \varphi:M_{h}\rightarrow{\mathbb{R}}\text{ }\left\vert \text{
}\varphi_{|T}\text{ is linear on each triangle }T\text{, and globally continuous}  \right.  \right\} \\
S_{h}^{\ast}  &  :=\left\{  \psi:M_{h}\rightarrow{\mathbb{R}}\text{
}\left\vert \text{ }\psi_{|T}\text{ is linear, and continuous at edge
midpoints}\right.  \right\}
\end{align*}

\begin{figure}[!h]
    \centering
    \includegraphics[width=.8\linewidth]{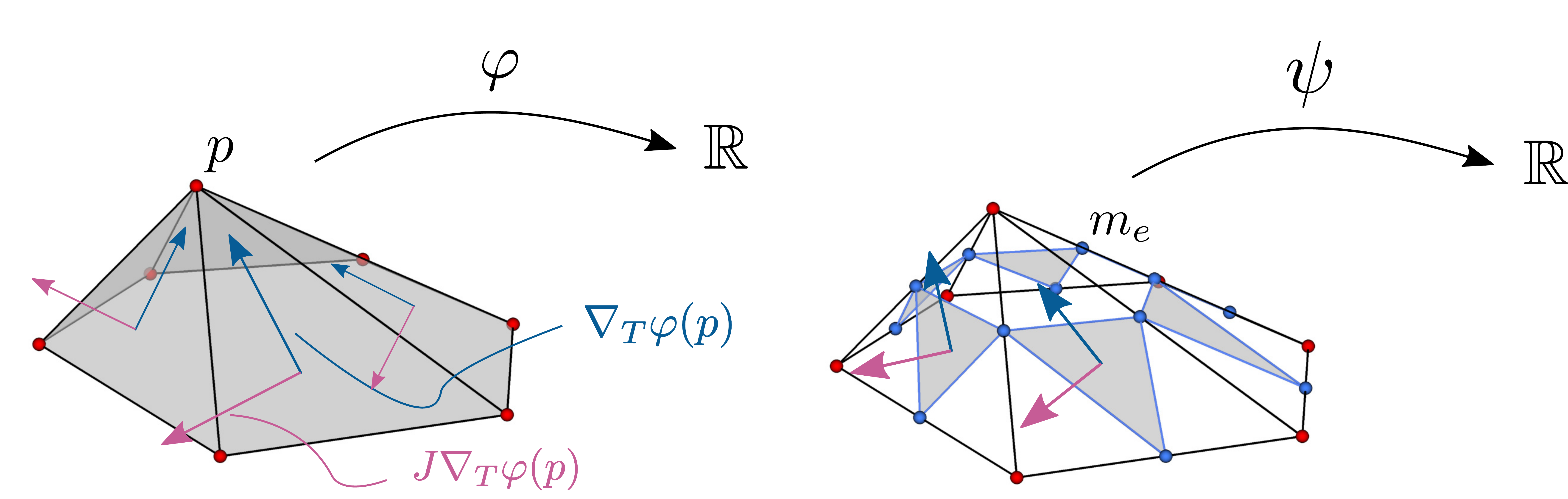}
    \caption{Examples of a function $\varphi \in S_h$ together with $\nabla \varphi$, $J\nabla\varphi$ (left) and $\psi \in S_h^{*}$ with $\nabla \psi$, and $J\nabla \psi$ (right) defined over the triangle $T$.}
    \label{fig:basis}
\end{figure}

A geometric realization of example functions on $S_h$ and $S^{*}_h$ is shown in figure~\ref{fig:basis}. 
Two additional subspaces $S_{0}\subset{S_{h}}$ and $S_{0}^{\ast}\subset{S_{h}^{\ast}}$ for surfaces with boundary are given by
\begin{align*}
S_{0}  &  :=\left\{  \varphi\in S_{h}\text{ }\vert \text{
}\varphi(v)=0\text{ for all boundary vertices }v\right\} \\
S_{0}^{\ast}  &  :=\left\{  \psi \in S_{h}^{\ast}\text{
}\vert\,\, \psi(m_e)=0\text{ for all boundary edge mid-points }m_e\right\}.
\end{align*}

The \emph{gradient field} $\nabla \varphi$ of a function $\varphi\in S_{h}$ or
$S_{h}^{\ast}$ is a constant tangent vector in each triangle. The
\emph{co-gradient field } $J\nabla \varphi$ is obtained by a rotation $J$ of the gradient $\nabla \varphi$ by $\frac{\pi}{2}$ in each triangle (see figure~\ref{fig:basis}). 
The idea of having functional spaces is a common technique in finite element analyses to solve complicated partial differential equations. 
For example a temperature map $u$ which assigns a scalar value to each vertex of $M_h$ is an element of $S_h$. It can be expressed with respect to the nodal basis functions $(\varphi_i)_i$ of $S_h$, i.e $u = \sum_{i} u_i \varphi_i$, where $\varphi_i$ is the Kronecker delta, $\varphi_i(v_j) = 1$ if $i=j$ and $0$ otherwise, for a vertex $v_j$ of $M_h$. Then, the gradient of an arbitrary function in $S_h$ can  simply be expressed as a linear combination of the $\nabla \varphi_i$'s.

\subsection{Vector fields on Simplicial Surfaces}
\begin{definition}[PCVF]
The space of \emph{piecewise constant tangential vector fields }$\Lambda
^{1}(M_{h})$ on a $2$-dimensional simplicial surface 
$M_{h}\subset {\mathbb{R}^{n}}$ is given by:
\[
\Lambda^{1}(M_{h}):=
\left\{  
\vfx:M_{h}\rightarrow TM_{h}\ \left\vert
\ \vfx_{|\text{triangle }T}
\text{ is a constant tangent vector in }T  
\right. \right\}.
\]
\end{definition}
Here, $TM_{h}$ denotes the (piecewise) tangent bundle of $M_h$. 
The gradient field $\nabla\varphi$ introduced previously is an example of a tangential vector field. 
\begin{definition}[$L^2$-\textit{product}]
The $L^2$-\textit{product} of two vector fields, $\vfx = (\vx_T)_{T\in M_h}$ and $\vfy=(\vy_T)_{T\in M_h}$, where $\vx_T,\vy_T$ are tangent vectors in the triangle $T$, is defined by the area-weighted Euclidean sum 
\[
    \left<\vfx,\vfy\right>_{L^2} = \sum_{T\in M_h} \left<\vx_T,\vy_T\right>\text{Area(T)}.
\]
\end{definition}

In particular, two vector fields $\vfx$ and $\vfy\in \vfx_h$ are \emph{$L^2$-orthogonal} if $\left<\vfx,\vfy\right>_{L^2}=0$. 
A vector field subspace $A\subseteq \Lambda
^{1}(M_{h})$ is the \emph{$L^2$-orthogonal decomposition} of two subspaces $B,C \subseteq A$ (written $A=B \oplus C$) if every $\vfx\in A$ can be written uniquely as a sum $\vfx = \vfy + \vfz$ with $\vfy\in B$ and $\vfz\in C$, and furthermore  $\left<\vfy,\vfz\right> = 0$. The sum of two vector fields is a new vector field obtained by the sum of the components.

\subsection{Discrete Calculus}
\begin{figure}
    \centering
    \includegraphics[width=\linewidth]{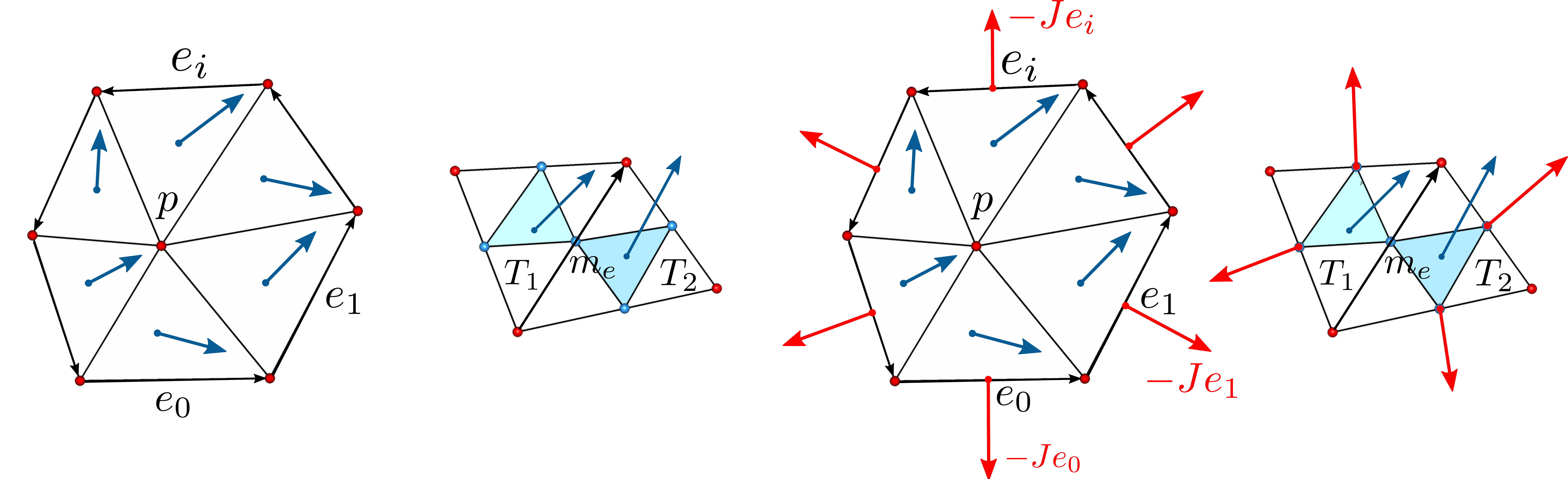}
    \caption{Computation of the curl and divergence of a vector field at a point on $\nabla S_h$ and $J\nabla S_h^{*}$.}
    \label{fig:rotdiv}
\end{figure}

\begin{definition}[Discrete Curl]
The \textit{discrete curl} of a vector field $\vfx=(\vx_T)_{T\in M_h}$ at a vertex $p$ and an edge midpoint $m_e$ of $M_h$ is computed by 
\begin{align*}
\operatorname{curl}_{h}\vfx(p)  & :=\frac{1}{2}\oint\limits_{\partial\operatorname{star}p}
\vfx=\frac{1}{2}\sum_{i=1}^{k}\left< \vx_{| T_i},e_{i}\right> \\
\operatorname{curl}_{h}^{\ast}\vx(m_e)  &  :={\oint\limits_{\partial\operatorname{star}e}}
\vfx=-\left<\vx_{|T_{1}},e\right> +\left<\vx_{|T_{2}},e\right>
\end{align*}
where the $e_{i}$'s are the edges of the oriented boundary of $\text{star }p$, the $T_i$'s the triangles adjacent to $p$ and $e$ the edge with midpoint $m_e$ (see figure~\ref{fig:rotdiv}).
\end{definition}

\begin{definition}[Discrete Divergence]The \textit{discrete divergence} of a vector field $\vfx$  at a vertex $p$ and an edge midpoint $m_e$ of $M_h$ is computed by

\begin{align*}
\operatorname{div}_{h}\vfx(p)  &  :=\frac{1}{2}%
{\displaystyle\oint\limits_{\partial\operatorname{star}p}}
\left\langle \vfx,\normal\right\rangle ds=-\frac{1}{2}\sum_{i=1}^{k}\left\langle
\vx_{|T_i},Je_{i}\right\rangle \\
\operatorname{div}_{h}^{\ast}\vfx(m_e)  &  :=%
{\displaystyle\oint\limits_{\partial\operatorname{star}e}}
\left\langle \vfx,\normal\right\rangle ds=\left\langle \vx_{|T_{1}},J_{|T_{1}%
}e\right\rangle +\left\langle \vx_{|T_{2}},J_{|T_{2}}e\right\rangle
\end{align*}
where $\normal$ is the outer unit normal along $\partial\operatorname{star}p$
resp. $\partial\operatorname{star}e$. Discrete rotation and divergence are related by $\operatorname{curl}%
_{h}J\vfx=\operatorname{div}_{h}\vfx$ and $\operatorname{curl}_{h}^{\ast
}J\vfx=\operatorname{div}_{h}^{\ast}\vfx$, compare Figure \ref{fig:rotdiv}.
\end{definition}
\begin{definition}[Dirichlet and Neumann field]
A vector field $\vfx$ is a \textit{Dirichlet field} (resp. \textit{Neumann field}) if $\vfx_{|T\in \partial M_h}$ is orthogonal (resp. ``almost'' parallel) to the boundary edge of $T$.  
\end{definition}
In the discrete case, the definition of Neumann fields is subtle due to technical properties of the chosen function spaces $S_h$ and $S_h^*$. 
They are not strictly parallel along the boundary as one might expect from the smooth case, but can deviate slightly. However, they are overall mostly parallel, so for simplicity one may imagine them as being just parallel, in perfect duality to the definition of Dirichlet fields. For technical details we refer the reader to \cite[Sec.~3.1]{Poelke2016126}.

The harmonic  Dirichlet field $\hd$ is for example a divergence-free and a curl-free field orthogonal to $\partial M_h$. Note that Dirichlet fields and Neumann fields do not exists on a closed surface. 
One can nevertheless define them using hard directional constraints on certain features of the underlying surface, e.g sharp features. 

\end{appendices}
\section{Statement}
\paragraph*{Ethics}The MRI data of CoA patients and volunteers were acquired in frames of a study, which was carried out according to the principles of the
Declaration of Helsinki and approved by the local Ethics Committee
at Charité-Universitätsmedizin Berlin. Written informed consent was obtained from
the participants and/or their guardians. The clinical study has been
registered with \verb|ClinicalTrials.gov|
\paragraph*{Data accessibility} There is not data deposition applicable for the paper.

\paragraph*{Funding}
This research was carried out in the framework of {MATHEON} supported by the Einstein Foundation Berlin within the ECMath Project CH18.
\paragraph*{Competing interests} The authors declare no competing interests.
\paragraph*{Authors' contributions}
F.H. Razafindrazaka, L.Goubergrits, and K. Polthier developed the project from conception to design. P. Yevtushenko was responsible in acquiring the patient models, reconstructing the geometry, and doing the numerical simulations. K. Poelke, and K. Polthier designed and implemented the Hodge decomposition algorithm. F.H. Razafindrazaka analysed the data using the method. L. Goubergrits, and P. Yevtushenko interpreted the results with .  F.H. Razafindrazaka, L. Goubergrits, P. Yevtushenko, and K. Poelke wrote the manuscript. All authors gave final approval for publication.
\bibliographystyle{vancouver}
\bibliography{literatur}

\begin{thebibliography}{10}

\bibitem{CHATZIZISIS20072379}
Chatzizisis YS, Coskun A, Jonas M, Edelman E, Feldman C, Stone P.
\newblock Role of Endothelial Shear Stress in the Natural History of Coronary
  Atherosclerosis and Vascular Remodeling: Molecular, Cellular, and Vascular
  Behavior.
\newblock Journal of the American College of Cardiology. 2007;49(25):2379 --
  2393.

\bibitem{Chiu009}
Chiu J, Chien S.
\newblock Effects of Disturbed Flow on Vascular Endothelium: Pathophysiological
  Basis and Clinical Perspectives.
\newblock Physiological Reviews. 2011;91(1):327--387.

\bibitem{Zhang2018}
Zhang B, Gu J, Qian M, Niu L, Ghista D.
\newblock Study of correlation between wall shear stress and elasticity in
  atherosclerotic carotid arteries.
\newblock BioMedical Engineering OnLine. 2018 Jan;17(1):5.

\bibitem{SOULIS2010867}
Soulis J, Fytanidis D, Papaioannou V, Giannoglou G.
\newblock Wall shear stress on LDL accumulation in human RCAs.
\newblock Medical Engineering \& Physics. 2010;32(8):867 -- 877.

\bibitem{Meng1924}
Meng H, Wang Z, Hoi Y, Gao L, Metaxa E, Swartz D, et~al.
\newblock Complex Hemodynamics at the Apex of an Arterial Bifurcation Induces
  Vascular Remodeling Resembling Cerebral Aneurysm Initiation.
  2007;38(6):1924--1931.

\bibitem{Boussel2008}
Boussel L, Rayz V, McCulloch C, Martin A, Acevedo-Bolton G, Lawton M, et~al.
\newblock Aneurysm Growth Occurs at Region of Low Wall Shear Stress
  Patient-Specific Correlation of Hemodynamics and Growth in a Longitudinal
  Study. 2008 08;39:2997--3002.

\bibitem{Stevens2017}
Stevens R, Grytsan A, Biasetti J, Roy J, Lindquist~Liljeqvist M, Gasser C.
\newblock Biomechanical changes during abdominal aortic aneurysm growth.
\newblock PLOS ONE. 2017 11;12(11):1--16.

\bibitem{Saikrishnan2015}
Saikrishnan N, Mirabella L, Yoganathan A.
\newblock Bicuspid aortic valves are associated with increased wall and
  turbulence shear stress levels compared to trileaflet aortic valves.
\newblock Biomechanics and Modeling in Mechanobiology. 2015 Jun;14(3):577--588.

\bibitem{Lozowy2017}
Lozowy R, Kuhn D, Ducas A, Boyd A.
\newblock The Relationship Between Pulsatile Flow Impingement and Intraluminal
  Thrombus Deposition in Abdominal Aortic Aneurysms.
\newblock Cardiovascular Engineering and Technology. 2017 Mar;8(1):57--69.

\bibitem{JING2016}
Jing L, Zhong J, Liu J, Yang X, Paliwal N, Meng H, et~al.
\newblock Hemodynamic Effect of Flow Diverter and Coils in Treatment of Large
  and Giant Intracranial Aneurysms.
\newblock World Neurosurgery. 2016;89:199 -- 207.

\bibitem{knobelsdorff2014}
von Knobelsdorff-Brenkenhoff F, Trauzeddel R, Barker A, Gruettner H, Markl M,
  Schulz-Menger J.
\newblock Blood flow characteristics in the ascending aorta after aortic valve
  replacement, a pilot study using 4D-flow MRI.
\newblock International Journal of Cardiology. 2014;170(3):426 -- 433.

\bibitem{ARZANI2018145}
Arzani A, Shadden S.
\newblock Wall shear stress fixed points in cardiovascular fluid mechanics.
\newblock Journal of Biomechanics. 2018;73:145 -- 152.

\bibitem{goobergrits2014}
Goubergrits L, Schaller J, Kertzscher U, Woelken T, Ringelstein M, Spuler A.
\newblock Hemodynamic impact of cerebral aneurysm endovascular treatment
  devices: coils and flow diverters.
\newblock Expert Review of Medical Devices. 2014;11(4):361--373.

\bibitem{Morrisheartjnl2015}
Morris P, Narracott A, von Tengg-Kobligk H, Silva~Soto D, Hsiao S, Lungu A,
  et~al.
\newblock Computational fluid dynamics modelling in cardiovascular medicine.
\newblock Heart. 2015;.

\bibitem{Rodriguez-Palomares2018}
Rodriguez-Palomares J, Dux-Santoy L, Guala A, Kale R, Maldonado G,
  Teixid{\'o}-Tur{\`a} G, et~al.
\newblock Aortic flow patterns and wall shear stress maps by 4D-flow
  cardiovascular magnetic resonance in the assessment of aortic dilatation in
  bicuspid aortic valve disease.
\newblock Journal of Cardiovascular Magnetic Resonance. 2018 Apr;20(1):28.

\bibitem{prakash2001}
Prakash S, Ethier C.
\newblock Requirements for Mesh Resolution in 3D Computational Hemodynamics.
  2001 05;123:134--44.

\bibitem{schwarz-1995}
Schwarz G.
\newblock Hodge decomposition: a method for solving boundary value problems.
\newblock Lecture notes in mathematics. Springer; 1995.

\bibitem{bhatia_survey-2013}
Bhatia H, Norgard G, Pascucci V, Bremer P.
\newblock The {H}elmholtz-{H}odge Decomposition - A Survey.
\newblock IEEE Transactions on Visualization and Computer Graphics.
  2013;19(8):1386--1404.

\bibitem{shonkwiler-2013}
Shonkwiler C.
\newblock Poincar{\'e} duality angles and the {D}irichlet-to-{N}eumann
  operator.
\newblock Inverse Problems. 2013;29(4).

\bibitem{Poelke2016126}
Poelke K, Polthier K.
\newblock Boundary-aware hodge decompositions for piecewise constant vector
  fields.
\newblock Computer-Aided Design. 2016;78:126 -- 136.
\newblock \{SPM\} 2016.

\bibitem{poelke2017}
Poelke K.
\newblock Hodge-Type Decompositions for Piecewise Constant Vector Fields on
  Simplicial Surfaces and Solids with Boundary.
\newblock Freie Universit\"{a}t Berlin; 2017.

\bibitem{polthier_preuss-2003}
Polthier K, Preuss E.
\newblock Identifying Vector Field Singularities using a Discrete {H}odge
  Decomposition.
\newblock In: Hege HC, Polthier K, editors. Visualization and Mathematics III.
  Springer Verlag; 2003. p. 113--134.

\bibitem{wardetzky2006thesis}
Wardetzky M.
\newblock Discrete Differential Operators on Polyhedral Surfaces - Convergence
  and Approximation.
\newblock Freie Universit{\"a}t Berlin; 2006.

\bibitem{azencot2015discrete}
Azencot O, Ovsjanikov M, Chazal F, Ben-Chen M.
\newblock Discrete Derivatives of Vector Fields on Surfaces--An Operator
  Approach.
\newblock ACM Transactions on Graphics (TOG). 2015;34(3):29.

\bibitem{Schaller2014}
Schaller J, Goubergrits L, Yevtushenko P, Kertzscher U, Riesenkampff E, Kuehne
  T.
\newblock Hemodynamic in Aortic Coarctation Using MRI-Based Inflow Condition.
\newblock Lecture Notes in Computer Science. 2014;8330:65 -- 73.

\bibitem{florian2018}
Hellmeier F, Nordmeyer S, Yevtushenko P, Bruening J, Berger F, Kuehne T, et~al.
\newblock Hemodynamic Evaluation of a Biological and Mechanical Aortic Valve
  Prosthesis Using Patient-Specific MRI-Based CFD.
\newblock Artificial Organs. 2017;42(1):49--57.

\bibitem{KARIMI201442}
Safoora K, Mahsa D, Paritosh V, Mitra D, Bahram D, Payman J.
\newblock Effect of rheological models on the hemodynamics within human aorta:
  CFD study on CT image-based geometry.
\newblock Journal of Non-Newtonian Fluid Mechanics. 2014;207:42 -- 52.

\bibitem{goubergrits2015}
Goubergrits L, Riesenkampff E, Yevtushenko P, Schaller J, Kertzscher U,
  Hennemuth A, et~al.
\newblock MRI-based computational fluid dynamics for diagnosis and treatment
  prediction: Clinical validation study in patients with coarctation of aorta.
\newblock Journal of Magnetic Resonance Imaging. 2015;41(4):909--916.

\bibitem{goubergrits2013}
Goubergrits L, Mevert R, Yevtushenko P, Schaller J, Kertzscher U, Meier S,
  et~al.
\newblock The impact of MRI-based inflow for the hemodynamic evaluation of
  aortic coarctation.
\newblock Annals of Biomedical Engineering. 2013;41:2575--2587.

\bibitem{ladisa2011}
LaDisa JF, Figueroa CA, Vignon-Clementel IE, Kim HJ, Xiao N, Ellwein LM, et~al.
\newblock Computational simulations for aortic coarctation: representative
  results from a sampling of patients.
\newblock Journal of Biomechanical Engineering. 2011;133:091008--1 -- 9.

\bibitem{vergara2017}
Vergara C, LeVan D, Quadrio M, Formaggia L, Domanin M.
\newblock Large eddy simulations of blood dynamics in abdominal aortic
  aneurysms.
\newblock Medical Engineering and Physics. 2017;47:38 -- 46.

\bibitem{miyazaki2017}
Miyazaki S, Itatani K, Furusawa T, Nishino T, Sugiyama M, Takehara Y, et~al.
\newblock Validation of numerical simulation methods in aortic arch using 4D
  Flow MRI.
\newblock Heart Vessels. 2017;32:1032 -- 1044.

\end{thebibliography}
\end{document}